\documentclass[showpacs,pra,aps]{revtex4}
\usepackage[dvips]{graphics,epsfig,color}
\begin{document}
\title{ Magnetic extraction of energy from accretion disc around \\a
rotating black hole\footnote{Supported by the National Natural
Science Foundation of China under Grant Nos 10173004, 10373006 and
10121503}}
\author{GONG Xiao-Long,  WANG Ding-Xiong\footnote{To whom
correspondence should be addressed.
 Email: dxwang@hust.edu.cn},   and YE Yong-Chun } \affiliation{Department
of Physics, Huazhong University of Science and Technology, Wuhan
430074 }

\author{April 18, 2004}

\begin{abstract}

An analytical expression for the disc power is derived based on an
equivalent circuit in black hole (BH) magnetosphere with a mapping
relation between the radial coordinate of the disc and that of
unknown astrophysical load. It turns out that this disc power is
comparable with two other disc powers derived in the Poynting flux
and hydrodynamic regimes, respectively. In addition, the relative
importance of the disc power relative to the BZ power is
discussed. It is shown that the BZ power is generally dominated by
the disc power except some extreme cases. Furthermore, we show
that the disc power derived in our model can be well fitted with
the jet power of M87.

\pacs{97.60.Lf, 98.62.Mw, 98.62.Js}
\end{abstract}

\maketitle

As is well known, jets exist in many astronomical cases, such as
active galactic nuclei, quasars, and young stellar objects. The
association of jets with magnetized accretion discs or magnetized
central objects (black holes or stars) is strongly supported by
the recent observations of Hubble Space Telescope (HST), Chandra,
and VLBI.$^{[1,2]}$ Different theoretical models have been
proposed for acceleration and collimation of jets, which can be
divided into two main regimes, the Poynting flux regime and the
hydromagnetic regime $^{[3 - 6]}$. Both regimes are related to a
poloidal magnetic field threading the disc, from which energy and
angular momentum are extracted. In the Poynting flux regime,
energy is extracted in Poynting flux (i.e. purely electromagnetic
energy), but in the form of magnetically driven material winds in
the latter regime.

Recently, Cao discussed the disc power in the hydrodynamic regime,
and found that the disc power (henceforth $P_{MHD} )$ is mainly
determined by the strength and the configuration of the poloidal
magnetic field.$^{[7]}$ Lee discussed the disc power in the
Poynting flux regime (henceforth $P_{EM}^I )$, and found that the
Poynting flux caused by a rotating magnetic field with Keplerian
angular velocity can balance the energy and angular momentum
conservation of a stationary accretion flow.$^{[8]}$ In this
Letter, we derive the disc power (henceforth $P_{EM}^{II} )$ by
using an equivalent circuit in black hole (BH) magnetosphere, and
compare $P_{EM}^{II} $ with $P_{MHD} $ and $P_{EM}^I $,
respectively. Throughout this Letter the geometric units $G = c =
1$ are used.

In order to derive the expression of the disc power $P_{EM}^{II} $ we make
the following assumptions:

1. The disc is thin, Keplerian, stable and perfectly conducting,
which lies in the equatorial plane of the BH with the inner
boundary being at the marginally stable orbit with the radius
$r_{in} = r_{ms} $.$^{[9]}$

2. The configuration of the poloidal magnetic field of the BH-disc
system is shown in Fig. 1. The poloidal magnetic field lines
thread both the BH horizon and the surrounding disc, connecting
the disc with the remote astrophysical load. The shape of each
field line threading the disc at radius $r$ has an analogy with
that in the asymptotic jet model.$^{[10]}$ It is assumed that the
asymptotic radius of each field line $r'$ is smaller than its
light radius $R_L $. The asymptotic radius $r'$ is related to the
disc radius $r$ by

\begin{equation}
\label{eq1}
{r}' = \lambda r
\end{equation}

3. The astrophysical load is axisymmetric, being located evenly in a
plane\textbf{\textit{ P}} with some height above the disc. The surface
resistivity $\sigma _L $of the load obeys the following relation, $\sigma _L
= \alpha _Z \sigma _H = 4\pi \alpha _Z $, where $\sigma _H = 4\pi =
377\mbox{ }ohm$ is the surface resistivity of the BH horizon, and $\alpha _Z
= 1$ is taken in calculations.

4. The poloidal magnetic field on the disc varies as a power law as given in
Refs.[11-13],

\begin{equation}
\label{eq2}
B_D^p \propto \xi ^{ - n},
\end{equation}

\noindent where $\xi \equiv r \mathord{\left/ {\vphantom {r
{r_{ms} }}} \right. \kern-\nulldelimiterspace} {r_{ms} }$ is the
radial parameter of the disc defined in terms of the radius of
inner edge of the disc. The field threading the BH is just a
continuation of the field threading the disc, and the following
relation is satisfied, $^{[7]}$

\begin{equation}
\label{eq3}
B_H^p = \varsigma B_D^p \left( {r_{ms} } \right),
\end{equation}

\noindent where $B_H^p $ and $B_D^p \left( {r_{ms} } \right)$ are
the magnetic fields at the horizon and the inner edge of the disc,
respectively. Some authors $^{[14,15]}$ argued that the magnetic
field threading the BH should not be significantly stronger than
the field threading the inner disc. In this Letter $\varsigma = 3$
is assumed in calculations.

\begin{figure}
\vspace{0.5cm}
\begin{center}
\includegraphics[width=7cm]{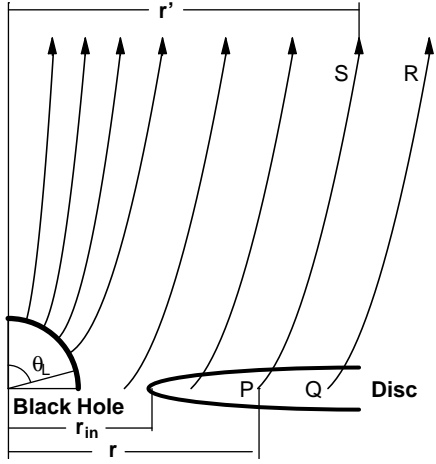}
\caption{Configuration of poloidal magnetic field threading the BH
and its surrounding disc.}
\end{center}
\end{figure}

\begin{figure}
\vspace{0.5cm}
\begin{center}
\includegraphics[width=8cm]{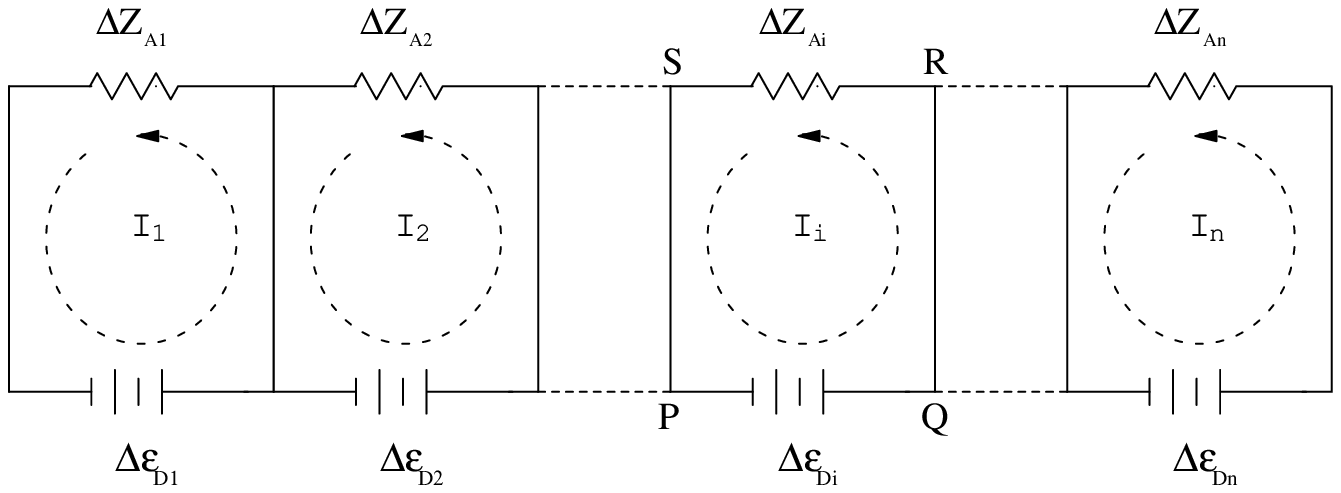}
\caption{Equivalent circuit for magnetic extraction of energy from
accretion disc around a rotating BH.}
\end{center}
\end{figure}

Based on MacDonald and Thorne's work,$^{[16]}$ we propose an equivalent
circuit to calculate the disc power as shown in Fig.2. The segments
\textbf{\textit{PS}} and \textbf{\textit{QR}} represent two adjacent
magnetic surfaces consisting of the field lines connecting the disc and the
load, and the segments \textbf{\textit{PQ}} and
\textbf{\textit{RS}} represent the disc surface and the load
sandwiched by the two surfaces, respectively. The quantities\textbf{
}$\Delta Z_A $ and $\Delta \varepsilon _D $ (the subscript ``$i$''
is omitted) are the resistance of the load and
\textbf{\textit{emf}} due to the rotation of the disc, respectively. The
disc load is neglected in calculations due to its perfect conductivity.

Based on assumptions (\ref{eq1}) and (\ref{eq2}) the load
resistance $\Delta Z_A $ between the two adjacent magnetic
surfaces can be written as

\begin{equation}
\label{eq4} \Delta Z_A = \sigma _L \frac{d{r}'}{2\pi {r}'} = 2
\frac{dr}{r}.
\end{equation}

\newpage
\noindent
The following equations are used in deriving disc power,
which are almost the same as given in deriving the BZ power and MC
power in our previous work. $^{[12,13]}$

\begin{equation}
\label{eq5}
\Delta P_{EM}^{II} = \left( {I^p} \right)^2\Delta Z_A
, \quad I^p = {\Delta \varepsilon _D } \mathord{\left/ {\vphantom
{{\Delta \varepsilon _D } {\Delta Z_A }}} \right.
\kern-\nulldelimiterspace} {\Delta Z_A }, \quad \Delta \varepsilon
_D = \left( {{\Delta \Psi _D } \mathord{\left/ {\vphantom {{\Delta
\Psi _D } {2\pi }}} \right. \kern-\nulldelimiterspace} {2\pi }}
\right)\Omega _D ,
\end{equation}

\noindent
where $I^p$ is the current in each loop, and $\Delta \Psi _D $ is the
magnetic flux between the two adjacent magnetic surfaces,

\begin{equation}
\label{eq6}
\Delta \Psi _D = B_D^p 2\pi \left( {{\varpi \rho }
\mathord{\left/ {\vphantom {{\varpi \rho } {\sqrt \Delta }}}
\right. \kern-\nulldelimiterspace} {\sqrt \Delta }}
\right)_{\theta = \pi \mathord{\left/ {\vphantom {\pi 2}} \right.
\kern-\nulldelimiterspace} 2} dr,
\end{equation}

\noindent The quantity $\Omega _D $ is the angular velocity of the
disc at the place where the magnetic flux penetrates, and it reads


\begin{equation}
\label{eq7} \Omega _D = \frac{1}{M(\xi ^{3 \mathord{\left/
{\vphantom {3 2}} \right. \kern-\nulldelimiterspace} 2}\chi
_{ms}^3 + a_ * )}.
\end{equation}

\noindent
The concerned Kerr metric coefficients are given as
follows,$^{[17]}$

\begin{equation}
\label{eq8}
\left\{ {\begin{array}{l}
 \varpi = \left( {\Sigma \mathord{\left/ {\vphantom {\Sigma \rho }} \right.
\kern-\nulldelimiterspace} \rho } \right)\sin \theta , \\
 \Sigma ^2 \equiv \left( {r^2 + a^2} \right)^2 - a^2\Delta \sin ^2\theta ,
\\
 \rho ^2 \equiv r^2 + a^2\cos ^2\theta , \\
 \Delta \equiv r^2 + a^2 - 2Mr. \\
 \end{array}} \right.
\end{equation}

\noindent Combining Eqs.(2) with (9), we have the poloidal
magnetic field on the disc expressed by

\begin{equation}
\label{eq9}
B_D^p = \varsigma ^{ - 1}B_H^p \xi ^{ - n}.
\end{equation}

\noindent Incorporating Eqs. (5)---(9), we have

\begin{equation}
\label{eq10}
{\Delta P_{EM}^{II} } \mathord{\left/ {\vphantom
{{\Delta P_{EM}^{II} } {P_0 }}} \right. \kern-\nulldelimiterspace}
{P_0 } = f\left( {a_ * ,\xi ,n} \right)d\xi ,
\end{equation}

\noindent
where the function $f(a_\ast ,\xi ,n)$ is expressed by

\begin{equation}
\label{eq11} f\left( {a_ * ,\xi ,n} \right) = \frac{\chi _{ms}^8
\left( {1 + \xi ^{ - 2}\chi _{ms}^{ - 4} a_ * ^2 + 2a_ * ^2 \xi ^{
- 3}\chi _{ms}^{ - 6} } \right)\xi ^{ - 2n + 3}}{2\alpha _Z
\varsigma ^2\left( {1 + \xi ^{ - 2}\chi _{ms}^{ - 4} a_ * ^2 -
2\xi ^{ - 1}\chi _{ms}^{ - 2} } \right)\left( {\xi ^{3
\mathord{\left/ {\vphantom {3 2}} \right.
\kern-\nulldelimiterspace} 2}\chi _{ms}^3 + a_ * } \right)^2}.
\end{equation}

\noindent
The quantity $P_0 $ is defined as

\begin{equation}
\label{eq12}
P_0 = B_H^2 M^2 \approx B_4^2 (M / M_ \odot )^2\times
6.59\times 10^{28}erg \cdot s^{ - 1},
\end{equation}

\noindent where $B_4 $ and $M \mathord{\left/ {\vphantom {M {M_
\odot }}} \right. \kern-\nulldelimiterspace} {M_ \odot }$ are the
strength of $B_H^p $ and the BH mass in the units of $10^4gauss$
and one solar mass, respectively. Integrating Eq.(10) over the
radial parameter $\xi $, we have the expression for the disc power
as follows,

\begin{equation}
\label{eq13}
{P_{EM}^{II} \left( {a_ * ,\xi ,n} \right)}
\mathord{\left/ {\vphantom {{P_{EM}^{II} \left( {a_ * ,\xi ,n}
\right)} {P_0 }}} \right. \kern-\nulldelimiterspace} {P_0 } =
\int_1^\xi f \left( {a_ * ,{\xi }',n} \right)d{\xi }',
\end{equation}

Cao derived the disc power in the hydromagnetic regime as
follows,$^{[7]}$

\newpage

\begin{equation}
\label{eq14}
L_d = \frac{1}{2}\int {\left( {B_D^p }
\right)^2\left[ {r\Omega _D \left( r \right)} \right]^\alpha }
f\left( {\alpha ,\gamma _j } \right)rdr,
\end{equation}

\begin{equation}
\label{eq15}
f\left( {\alpha ,\gamma _j } \right) = \left( {\gamma
_j - 1} \right)\gamma _j^\alpha \left( {\gamma _j^2 - 1} \right)^{
- {\left( {\alpha + 1} \right)} \mathord{\left/ {\vphantom
{{\left( {\alpha + 1} \right)} 2}} \right.
\kern-\nulldelimiterspace} 2},
\end{equation}

\noindent
where the function $f\left( {\alpha ,\gamma _j } \right)$ depends on the
self-similar index $\alpha > 1$ and the Lorentz factor of the jet $\gamma _j
$. Since both $r\Omega _D \left( r \right)$ and $f\left( {\alpha ,\gamma _j
} \right)$ are always less than unity, the upper limit to the disc power
$L_d $ can be written as

\begin{equation}
\label{eq16}
L_d < P_{MHD} = \frac{1}{2}\int {\left( {B_D^p }
\right)^2r^2\Omega _D \left( r \right)dr} .
\end{equation}

\noindent
The expression of the poloidal magnetic field is
expressed as follows,$^{[8]}$

\begin{equation}
\label{eq17}
B_D^p \left( r \right) \propto \left( {r
\mathord{\left/ {\vphantom {r M}} \right.
\kern-\nulldelimiterspace} M} \right)^{{ - 3} \mathord{\left/
{\vphantom {{ - 3} 4}} \right. \kern-\nulldelimiterspace} 4}\left(
{M \mathord{\left/ {\vphantom {M {M_ \odot }}} \right.
\kern-\nulldelimiterspace} {M_ \odot }} \right)^{{ - 1}
\mathord{\left/ {\vphantom {{ - 1} 2}} \right.
\kern-\nulldelimiterspace} 2}A^{ - 1}BE^{1 \mathord{\left/
{\vphantom {1 2}} \right. \kern-\nulldelimiterspace} 2},
\end{equation}

\noindent where $A$, $B$ and $E$ are general relativistic
correction factors given in Ref.[9].Combining Eq.(16) with
Eq.(17), we obtain

\begin{equation}
\label{eq18}
{P_{MHD} \left( {a_ * ,\xi } \right)} \mathord{\left/
{\vphantom {{P_{MHD} \left( {a_ * ,\xi } \right)} {P_0 }}} \right.
\kern-\nulldelimiterspace} {P_0 } = \frac{1}{2\varsigma
^2}\int_1^\xi {\frac{\chi _{ms}^3 {\xi }'^{ - 1}}{{\rm K}^2\left(
{1 + {\xi }'^{{ - 3} \mathord{\left/ {\vphantom {{ - 3} 2}}
\right. \kern-\nulldelimiterspace} 2}\chi _{ms}^{ - 3} a_ * }
\right)}} d{\xi }',
\end{equation}

\noindent
where ${\rm K} = {\left( {A^{ - 1}BE^{1 \mathord{\left/ {\vphantom {1 2}}
\right. \kern-\nulldelimiterspace} 2}} \right)_{r_{in} } } \mathord{\left/
{\vphantom {{\left( {A^{ - 1}BE^{1 \mathord{\left/ {\vphantom {1 2}} \right.
\kern-\nulldelimiterspace} 2}} \right)_{r_{in} } } {\left( {A^{ - 1}BE^{1
\mathord{\left/ {\vphantom {1 2}} \right. \kern-\nulldelimiterspace} 2}}
\right)_r }}} \right. \kern-\nulldelimiterspace} {\left( {A^{ - 1}BE^{1
\mathord{\left/ {\vphantom {1 2}} \right. \kern-\nulldelimiterspace} 2}}
\right)_r }$.

Lee derived the disc power $P_{EM}^I $ in the Poynting flux regime
as follows,$^{ [8]}$

\begin{equation}
\label{eq19} P_{EM}^I = 4\left\{ {\left[ { - u_0 \left( {r_{out} }
\right)} \right] - \left[ { - u_0 \left( {r_{in} } \right)}
\right]} \right\}\left[ {B_D^p \left( {r_{in} } \right)r_{in} }
\right]^2.
\end{equation}

\noindent
Setting the inner and outer edge of the disc at $r_{in}
= r_{ms} $ and $r_{out} = \xi r_{ms} $, we have

\begin{equation}
\label{eq20} {P_{EM}^I \left( {a_ * ,\xi } \right)}
\mathord{\left/ {\vphantom {{P_{EM}^I \left( {a_ * ,\xi } \right)}
{P_0 }}} \right. \kern-\nulldelimiterspace} {P_0 } = 4\varsigma ^{
- 2}\chi _{ms}^4 \left( {E^\dag - E_{ms}^\dag } \right),
\end{equation}

\noindent where $E^\dag = - u_0 \left( {\xi r_{ms} } \right)$ and
$E_{ms}^\dag = - u_0 \left( {r_{ms} } \right)$ are the specific
energies of the disc matter corresponding to $r$ and $r_{ms} $,
respectively.$^{ [9]}$

We are going to compare our disc power $P_{EM}^{II} $ with
$P_{MHD} $ and $P_{EM}^I $. To facilitate comparison between
$P_{EM}^{II} $ and $P_{MHD} $, we take $n = 3 \mathord{\left/
{\vphantom {3 4}} \right. \kern-\nulldelimiterspace} 4$ in
Eq.(13), which is the same as given in Eq.(17). By using Eqs.(13),
(18) and (20) we have the curves of the three disc powers varying
with the radial parameter $\xi $ for the given values of $a_ * $
as shown in Fig. 3.

\begin{figure*}
\begin{center}
{\includegraphics[width=5.0cm]{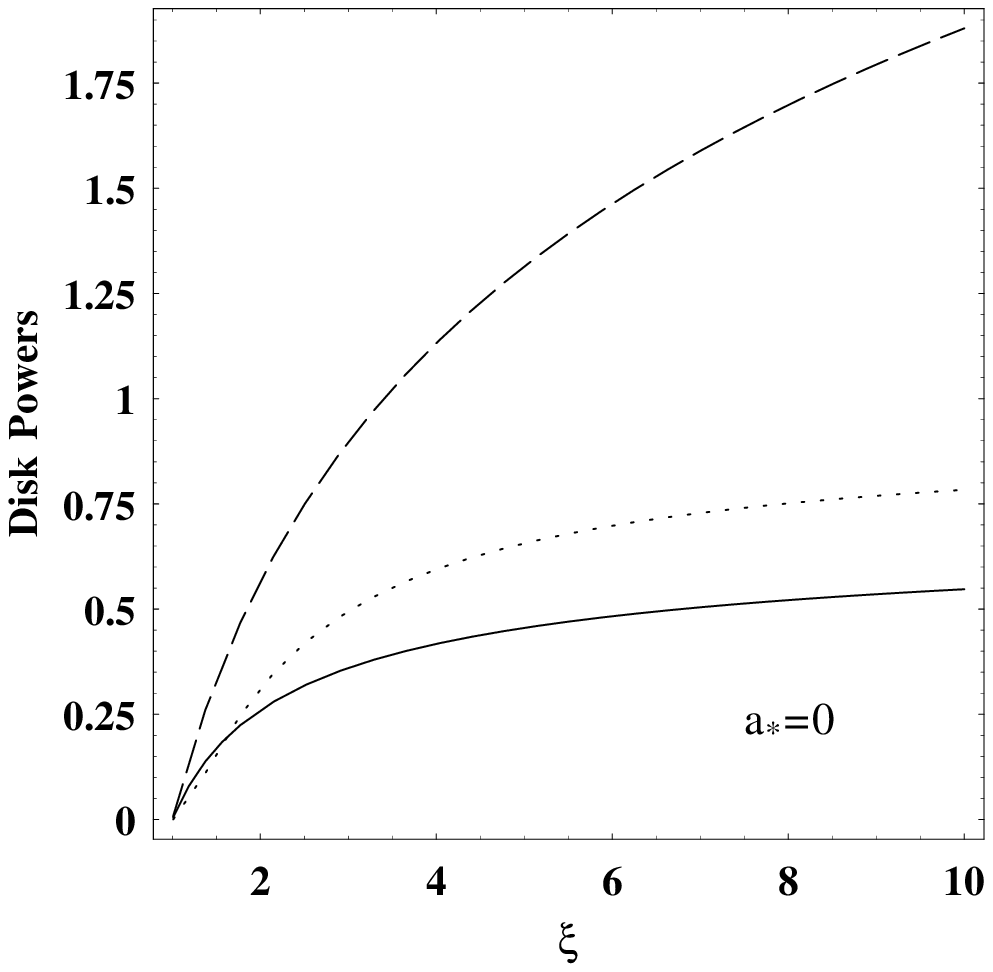} \hfill
\includegraphics[width=5.0cm]{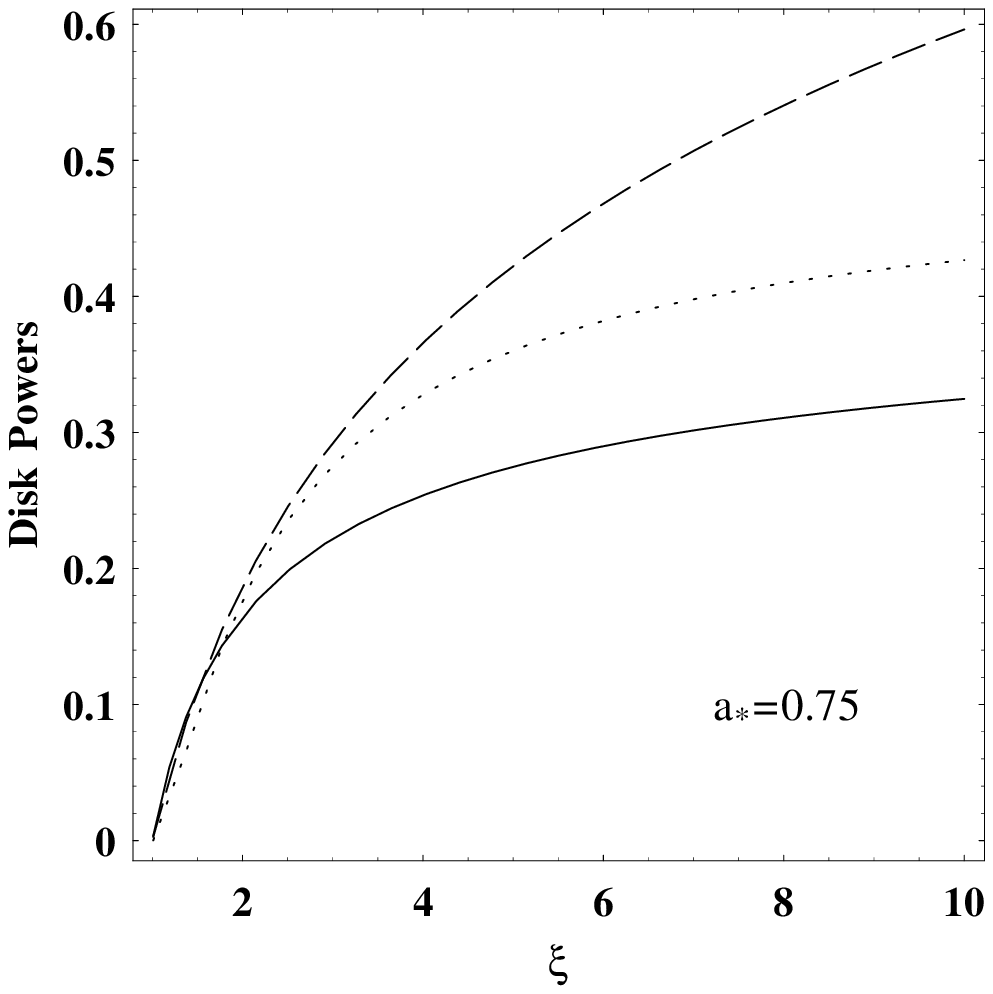} \hfill
\includegraphics[width=5.0cm]{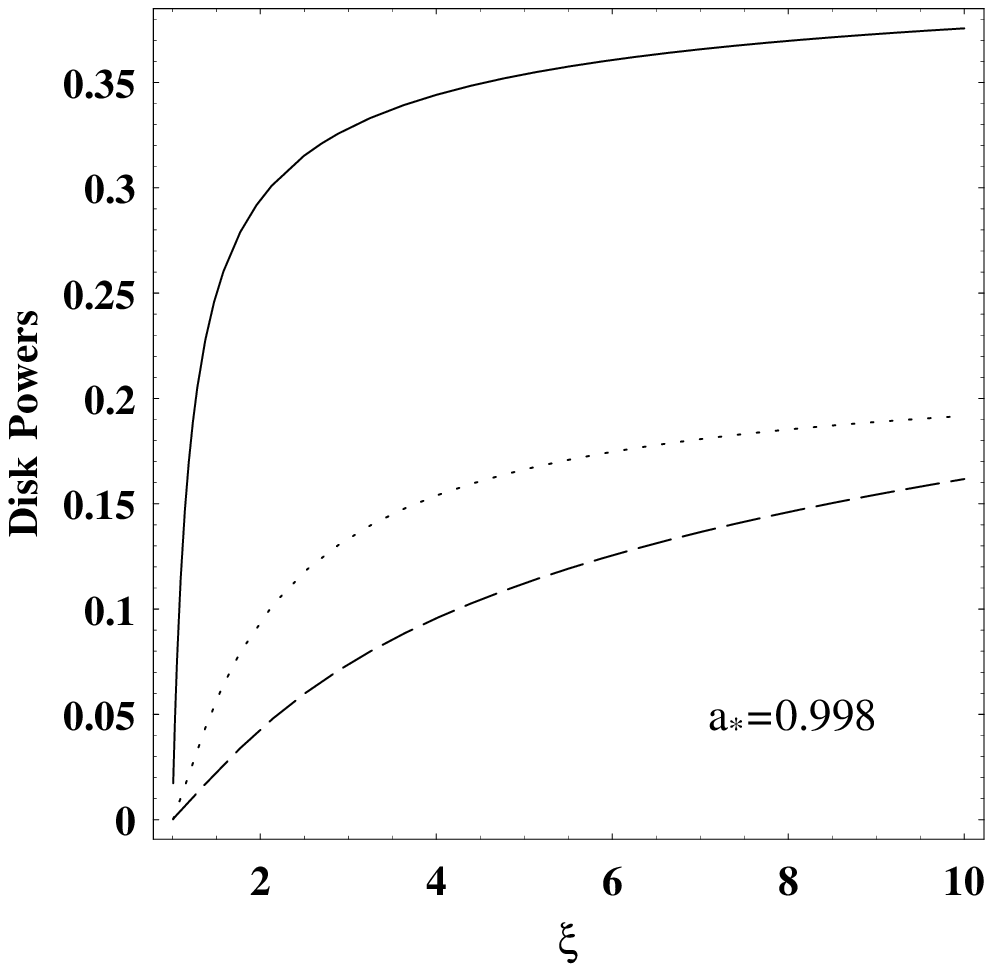}
\centerline{\hspace{0.6cm}(a)\hspace{5.65cm}(b)\hspace{5.7cm}(c)}
} \caption{The curves of $P_{EM}^{II} $ (solid line), $P_{MHD} $
(dashed line) and $P_{EM}^I $(dotted line) versus $\xi $ for $1 <
\xi < 10$ with different values of $a_
* $.}\label{fig3}
\end{center}
\end{figure*}

In order to compare the above disc powers more clearly, we define
$\eta_{1} \equiv P_{EM}^{II}/P_{MHD}$ and $\eta_{2} \equiv
P_{EM}^{II}/P_{EM}^{I}$, and have the contours of $\eta _1 $ and
$\eta _2 $ with different values in $\xi - a_ * $ parameter spaces
as shown in Fig. 4 (a) and (b), respectively.

\begin{figure*}
\begin{center}
{\includegraphics[width=5.5cm]{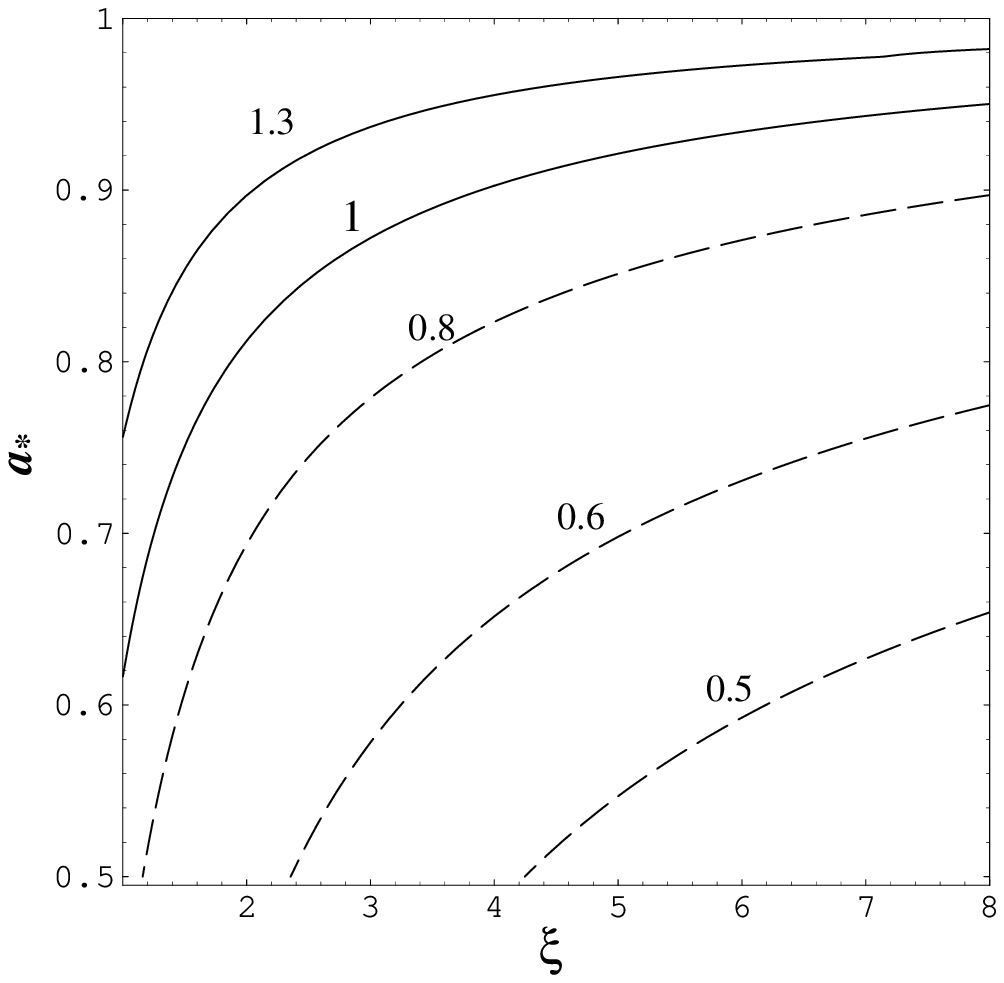}\hspace{1.cm}
\includegraphics[width=5.5cm]{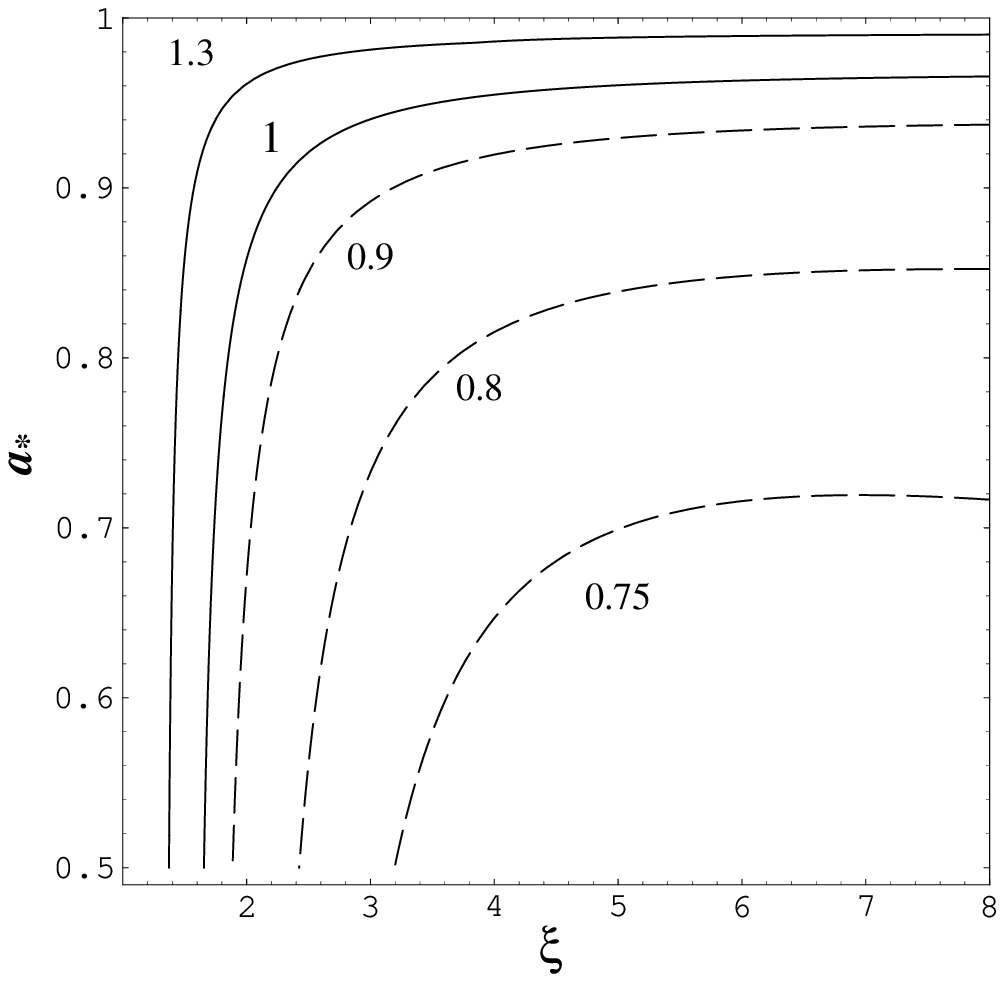}
\centerline{\hspace{1.0cm}(a)\hspace{5.65cm}(b)} }
 \caption{The contours of the ratios of the disc powers in $a_\ast - \xi $
parameter space, (a) $\eta_{1} \equiv P_{EM}^{II}/P_{MHD}$, and
(b) $\eta_{2} \equiv P_{EM}^{II}/P_{EM}^{I}$.}
\end{center}
\end{figure*}

Inspecting Figs.3 and 4, we have the following results:

(1) If the BH spin is not extremely high, the three disc powers satisfy the
relation, $P_{EM}^{II} < P_{EM}^I < P_{MHD} $. However this order reverses
to $P_{MHD} < P_{EM}^I < P_{EM}^{II} $, when the BH spin approaches unity,

(2) The disc power $P_{EM}^{II} $ dominates over $P_{EM}^I $ and
$P_{MHD} $ in two cases: (i) The region is very close to the inner
edge of the disc, (ii) The BH spin is very high.

Recently some authors argued that the BZ power was overestimated, and it is,
in general, dominated by the electromagnetic power output of the inner
region of the disc, provided that the poloidal magnetic field threading the
BH does not differ significantly in strength from that threading the
disc.$^{[14, 15]}$ In this Letter we shall discuss the importance of the
disc power relative to the BZ power. The BZ power has been derived also
based an equivalent circuit in BH magnetosphere in our previous work.$^{
[12,13]}$

\begin{equation}
\label{eq21}
{P_{BZ} } \mathord{\left/ {\vphantom {{P_{BZ} } {P_0
}}} \right. \kern-\nulldelimiterspace} {P_0 } = 2a_ * ^2
\int_0^{\theta _L } {\frac{k\left( {1 - k} \right)\sin ^3\theta
d\theta }{2 - \left( {1 - q} \right)\sin ^2\theta }} ,
\end{equation}

\newpage

\noindent where $q \equiv \sqrt {1 - a_ * ^2 } $ is a function of
the BH spin, and $k \equiv {\Omega _F } \mathord{\left/ {\vphantom
{{\Omega _F } {\Omega _H }}} \right. \kern-\nulldelimiterspace}
{\Omega _H }$ is the ratio of the angular velocity of the magnetic
field lines to that of the horizon. The angular coordinate $\theta
$ varies from $0$ to $\theta _L $ on the BH horizon as shown in
Fig.1. In calculations $k = 0.5$ and $\theta _L = 0.45\pi $ are
assumed in Eq.(21). In order to compare the disc power with the BZ
power clearly, we defined the ratio of the disc power to the BZ
power as $\eta = {P_{EM}^{II} } \mathord{\left/ {\vphantom
{{P_{EM}^{II} } {P_{BZ} }}} \right. \kern-\nulldelimiterspace}
{P_{BZ} }$. For the given values of power-law index $n$ the
contours of $\eta $ are shown in $\xi - a_\ast $ parameter space
in Fig.5.

\begin{figure*}
\begin{center}
{\includegraphics[width=5.cm]{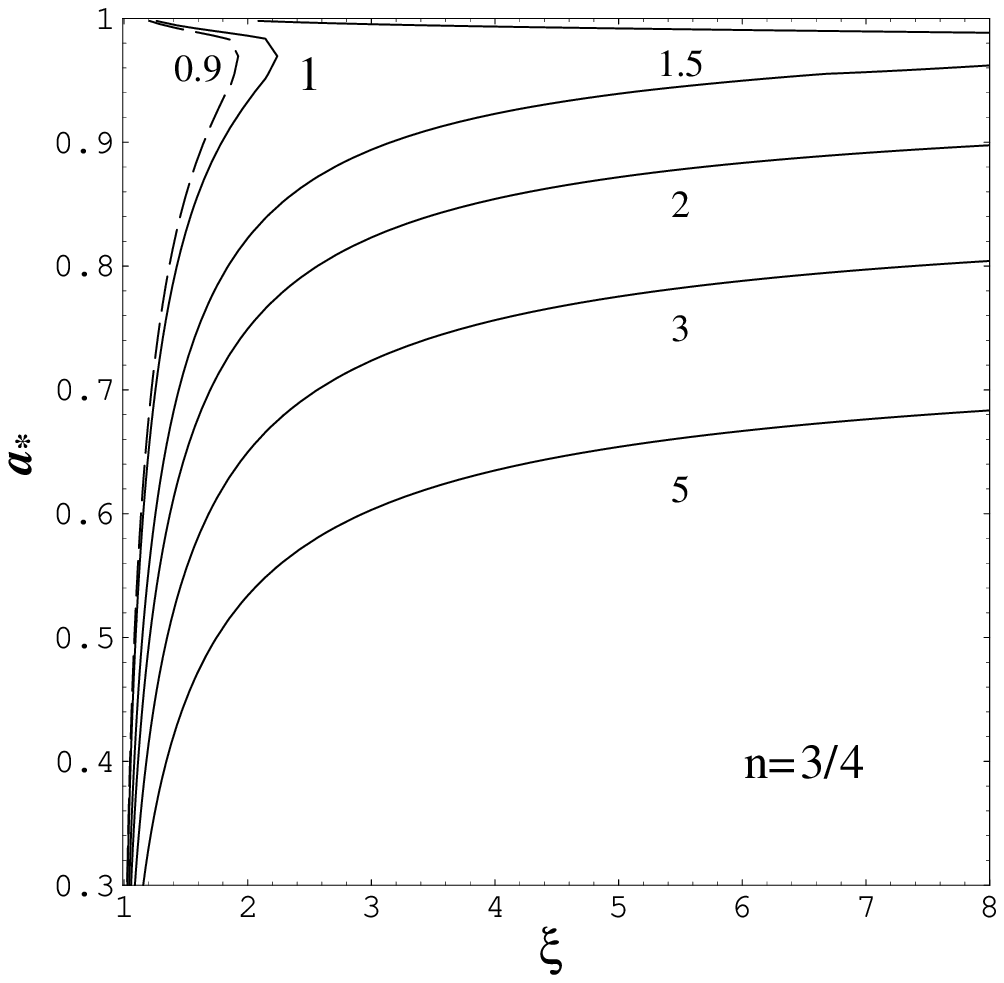} \hfill
\includegraphics[width=5.cm]{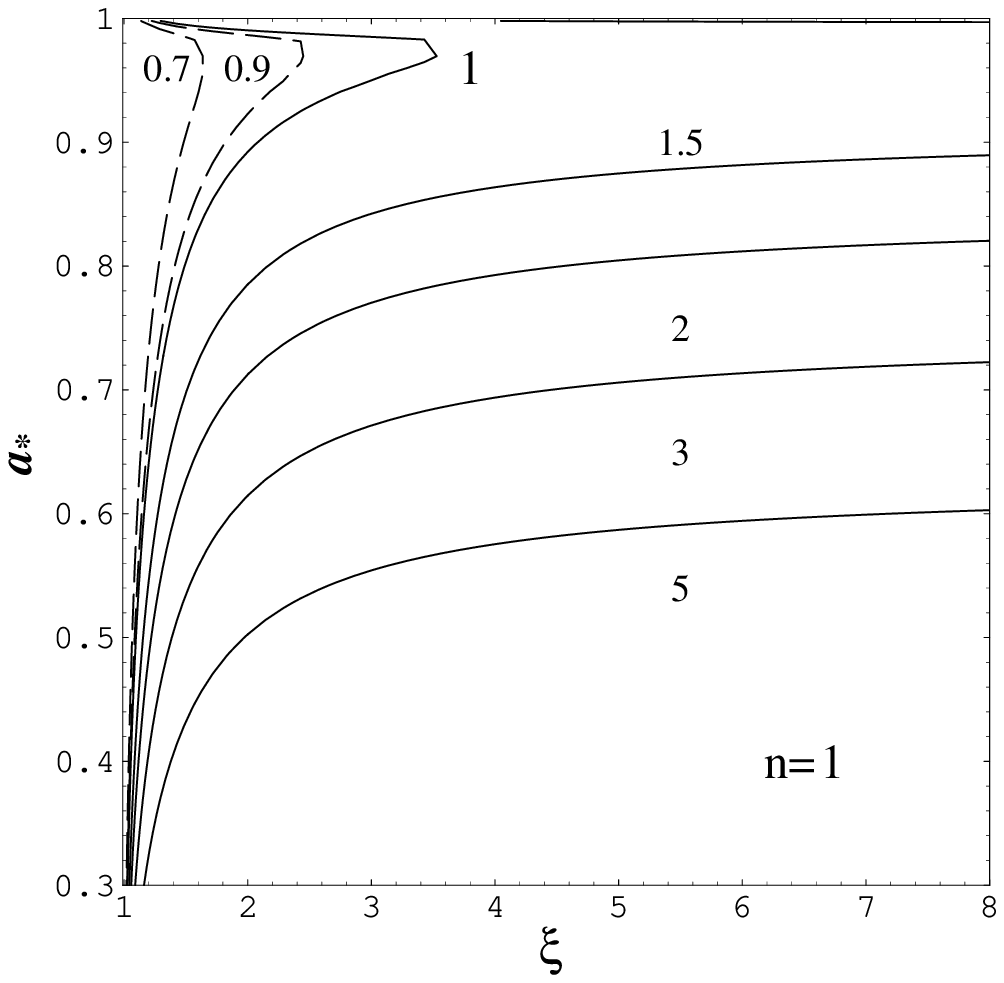} \hfill
\includegraphics[width=5.cm]{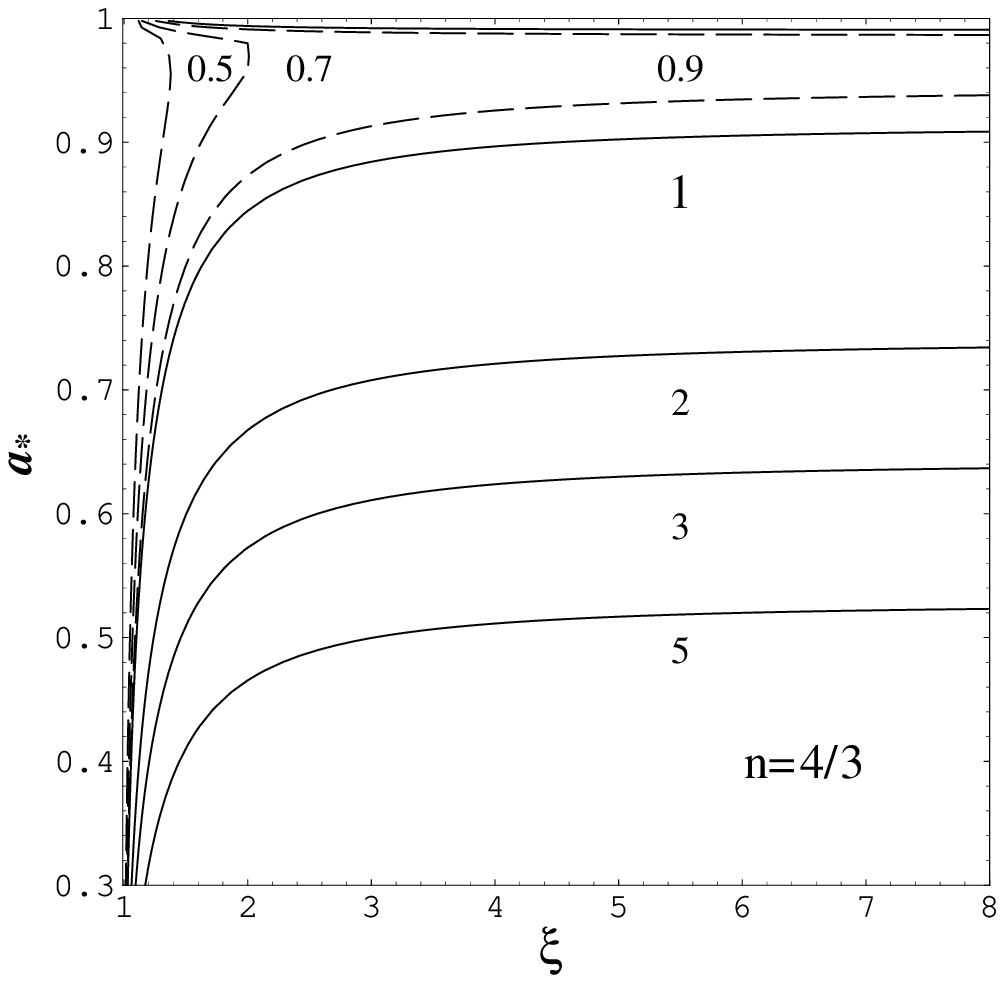}
\centerline{\hspace{0.6cm}(a)\hspace{5.65cm}(b)\hspace{5.7cm}(c)}
} \caption{The contours of constant values of $\eta $ in $\xi -
a_\ast $ parameter space for $n=3/4, n=1, n=4/3 $ in (a), (b) and
(c), respectively.}
\end{center}
\end{figure*}

From Fig.5 we find that the disc power will, in general, dominate over the
BZ power, except that the BH spin $a_\ast $ is extremely high. This result
is consistent with those given in Refs.[14] and [15].

In our model, different magnetic flux surfaces anchor at different foot
point radii $r$ on the disc, which correspond to different light cylinder
radii $R_L $. It is reasonable that the asymptotic radius of the magnetic
flux surface is not greater than its light cylinder radius, i.e., ${r}' \le
R_L $. The light cylinder radius satisfy the condition,

\begin{equation}
\label{eq22}
 R_L \Omega _D = 1.
\end{equation}

\noindent Incorporating Eqs.(7) and (22), we obtain the location
of the light cylinder radius of magnetic flux surface as follows,

\begin{equation}
\label{eq23}
 R_L = M\left[ {\left( {\sqrt \xi \chi _{ms} }
\right)^3 + a_ * } \right].
\end{equation}

Fendt and Memola defined jet expansion rate as the ratio of the
asymptotic jet radius to the foot point jet radius (the `disc
radius'), and the structure of jet is well in accord with the
observation of the M87 jet.$^{[10]}$ The observations of the HST
show that the radius of the M87 jet is about $100R_S $,$^{[18]}$
where $R_S = {2GM} \mathord{\left/ {\vphantom {{2GM} {c^2}}}
\right. \kern-\nulldelimiterspace} {c^2}$ is Schwarzschild radius.
Taking the jet expansion rate $\lambda = 10$ as given in
Ref.[10],we infer that the foot point jet radius is no greater
than $10R_S $.

By using Eq.(13), we have the disc power corresponding to
different BH spin $a_\ast $ as listed in Table 1, where $\xi $ is
the upper limit of integration in Eq.(13). In calculations the
mass of the central BH is $3\times 10^9M_ \odot $ as given in
Refs.[1] and [2], and the magnetic field on the BH horizon is
assumed to be about $3\times 10^2gauss$ as estimated in Ref.[3].
It turns out that the values of the disc power $P_{EM}^{II} $ in
Table 1 are consistent with the value of the M87 jet
power.$^{[19]}$

In this Letter the interaction between the BH and the surrounding disc is
neglected for simplification. In fact the magnetic coupling$^{[12,13,20]
}$of the BH with its surrounding disc will result in a further decrease of
the BZ power, and the domination of the disc power over the BZ power will be
strengthened in this case. We shall discuss this situation in our future
work.

\begin{table}[htbp]
\caption{Values of $P_{EM}^{II} $ in accord with the M87 jet power
for different values of the BH spin $a_\ast $.}\label{tab1}
\begin{tabular}
{p{77pt}p{69pt}p{69pt}p{69pt}p{69pt}p{69pt}} \hline\hline
 $a_{\ast}$& 0& 0.3& 0.5& 0.75&
0.998 \\
\hline ${r_{ms} } \mathord{\left/ {\vphantom {{r_{ms} } {R_S }}}
\right. \kern-\nulldelimiterspace} {R_S }$& 3& 2.489& 2.117&
1.579&
0.618 \\
\hline $\xi = {10R_S } \mathord{\left/ {\vphantom {{10R_S }
{r_{ms} }}} \right. \kern-\nulldelimiterspace} {r_{ms} }$& 3.333&
4.017& 4.725& 6.333&
16.169 \\
\hline $P_{EM}^{II} (erg \cdot s^{ - 1})$& $2.04\times 10^{44}$&
$1.91\times 10^{44}$& $1.78\times 10^{44}$& $1.57\times 10^{44}$&
$2.06\times 10^{44}$ \\
\hline
\end{tabular}
\end{table}

\noindent\textbf{References }

[1] Ford H C et al.1994 \textit{Astrophys. J.} \textbf{\textit{435}} L27

[2] Harms R J et al.1994 \textit{Astrophys. J.}\textbf{\textit{ 435}} L35

[3] Blandford R D and Znajek R L 1977 \textit{Mon. Not. R. Astron. Soc.} \textbf{\textit{179}} 433

[4] Lovelace R V E 1976 Nature \textbf{\textit{262}} 649

[5] Blandford R D and Payne D G 1982 \textit{Mon. Not. R. Astron. Soc.} \textbf{\textit{199}} 883

[6] Lovelace R V E et al. 2002 \textit{Astrophys. J.} \textbf{572 }L445

[7] Cao X W 2002 \textit{Mon. Not. R. Astron. Soc.}
\textbf{\textit{332}} 999

[8] Lee H K 2001 Phys. Rev. \textbf{\textit{D64}} 043006

[9] Novikov I D and Thorne K S 1973 in \textit{Black Holes}, ed.
Dewitt C Gordon and Breach, New York.

[10] Fendt C and Memola E 2001 Astron.{\&}Astrophys.\textbf{\textit{ 365}} 631

[11] Blandford R D 1976\textit{Mon. Not. R.
Astron.Soc.}\textbf{\textit{176}} 465

[12] Wang D X, Xiao K and Lei W H 2002 \textit{Mon. Not. R. Astron. Soc.} \textbf{\textit{335}} 655

[13] Wang D X Lei W H and Ma R Y 2003 \textit{Mon. Not. R. Astron. Soc.}\textbf{\textit{ 342}} 851

[14] Livio M, Ogilvie G L and Pringle J E 1999\textit{ Astrophys. J.} \textbf{\textit{512 }}100

[15] Ghosh P and Abramowicz M A 1997\textit{ Mon. Not. R. Astron. Soc.} \textbf{\textit{292}} 887

[16] MacDonald D and Thorne K S 1982 \textit{Mon. Not. R. Astron. Soc.} \textbf{\textit{198}} 345

[17] Thorne K S, Price R H and Macdonald D A 1986 \textit{Black Holes: The Membrane Paradigm},

\quad\quad Yale Univ. Press, New Haven

[18] Tsvetanov Z I et al. 1998 \textit{Astrophys. J. }\textbf{\textit{493}} L83.

[19] Bicknell G V and Begelman M C 1996 \textit{Astrophys. J.} \textbf{\textit{467}} 597

[20] Wang D X, Ma R Y, Lei W H and Yao G Z 2003 \textit{Astrophys. J.} \textbf{595} 109

\end{document}